# Local viscoelasticity of living cells measured by rotational magnetic spectroscopy

**J.-F. Berret**
*Matière et Systèmes Complexes, UMR 7057 CNRS Université Denis Diderot Paris-VII, Bâtiment Condorcet, 10 rue Alice Domon et Léonie Duquet, 75205 Paris, France.*

**Abstract:** When submitted to a magnetic field, micron size wires with superparamagnetic properties behave as embedded rheometers and represent interesting sensors for micro-rheology. Here we use rotational magnetic spectroscopy to measure the shear viscosity of the cytoplasm of living cells. We address the question of whether the cytoplasm is a viscoelastic liquid or a gel. The main result of the study is the observation of a rotational instability between a synchronous and an asynchronous regime of rotation, found for murine fibroblasts and human cancer cells. For wires of susceptibility 3.6, the transition occurs in the range 0.01 – 1 rad s$^{-1}$. The determination of the shear viscosity (10 – 100 Pa s) and elastic modulus (5 – 20 Pa) confirms the viscoelastic character of the cytoplasm. In contrast to earlier studies, it is concluded that the interior of living cells can be described as a viscoelastic liquid, and not as an elastic gel.

In rheology, viscoelastic liquids and viscoelastic solids differ from each other by their stress relaxation function $G(t)$. In viscoelastic liquids, $G(t)$ is a decreasing function of the time which tends to zero as $t \to \infty$, whereas in viscoelastic solids $G(t)$ tends to a finite elastic modulus, indicating the presence of residual unrelaxed stresses even at very long time[1]. Hence, the static shear viscosity, which is defined as $\eta_0 = \int_0^\infty G(t)\, dt$ takes a finite value for liquids and it is generally expressed as a product of a relaxation time and of an elastic modulus. For viscoelastic solids, the static viscosity is not defined, and the rheological properties are described in terms of a low frequency elastic modulus $G'(\omega \to 0)$ and of a yield stress, *i.e.* the critical value of the stress above which the sample does flow. In the following, viscoelastic solids will be termed "gels" for convenience. Here we address the question of the rheological properties of the intracellular medium of mammalian cells, and in particular the question to know whether the cytoplasm of such cells is a viscoelastic liquid or a gel.

Rheometers are used to determine the relationship between the strain and stress on material sample of volume of the order of milliliters. Microrheology in contrast uses micron-size probes embedded in the material and needs much less sample, of the order of 1 picoliter. The past 20 years have seen increasingly rapid advances in this field, specifically in cell and tissue biomechanics[2,3]. For living cells, techniques including Atomic Force Microscopy (AFM)[4,5], optical and magnetic tweezers[6-10], parallel microplates[11] and active tracking of internalized particles[7,10,12-14] were developed and provide consistent evaluation of the time or frequency dependencies of rheological parameters, such as the elastic compliance $J(t)$ or of the complex modulus $G^*(\omega)$. Numerous studies performed on a wide variety of cell lines have shown that $J(t)$[9,14,15] and $G^*(\omega)$[8,9,12,14,16] obey scaling laws as a function of time or frequency, with exponents comprised between 0.1 and 0.5. As for $G^*(\omega)$ measurements, the elastic modulus $G'(\omega)$ was found to be larger than the loss modulus $G''(\omega)$ on broad frequency ranges. These results were interpreted as an indication that the mechanical response of the cell or of the cytoplasm is that of a weak elastic solid[8-10,12,15,17], or in the terminology adopted previously that of a gel. Out-of-equilibrium modeling suggests that the cell interior is a kind of soft glassy material with an effective temperature higher than that of the physiological temperature[17]. Alternative approaches of biomechanics were also attempted and provided different viewpoints. Based on AFM micro-indentation and force-relaxation tests, Moeendarbary *et al.* proposed that the cytoplasm behaves as a poroelastic material, where the water redistribution within the cytoplasm plays a fundamental role in setting the internal rheology[5]. Other studies, in particular the earlier work of the Sackmann group[6,7,18] suggest that the intracellular medium behaves like a liquid and is characterized by a finite value of the shear viscosity[13]. More recently, Kuimova *et al.* developed porphyrin-dimer-based molecular rotors which fluorescence emitting frequencies scale with the viscosity of the surrounding medium. Evaluated on immortalized cervical cancer cells (HeLa cells), these authors found intracellular viscosities in the range 0.05 – 0.2 Pa s[19,20]. Chevry *et al.* show that micron-size wires internalized in mouse fibroblasts and MDCK epithelial cells exhibit Brownian-like orientational fluctuations, associated to finite effective viscosity of 0.1 - 1 Pa s, in relative agreement with the Kuimova *et al.*'s determinations[21].

Regarding the complexity of the dynamics in living systems, and the contrasting results obtained so far there is a need to develop new types of probes and protocols for testing the cell biomechanics. Following the pioneer work by Crick and Hughes some 60 years ago[22], recent studies have shown that microrheology based of the tracking and monitoring of anisotropic probes could bring significant advances to the field[23-34]. It has been proposed, for example, that the shear viscosity of a fluid could be determined from the motion of a micro-actuator submitted to a rotating electric[28] or magnetic field[27,32-36]. These techniques are described as electric or magnetic rotational spectrometry[29,30]. With micro-actuators rotating at increasing frequency, a transition between a steady and a hindered motion is predicted for viscous fluids, and it is found experimentally[25,27,29,30,37,38]. A recent quantitative analysis shows that for viscoelastic liquids, the elasticity does not affect the onset of the instability and that the expression for the critical frequency remains the same as in a viscous fluid[25]. For a gel of infinite static viscosity, in contrast, extrapolation yields a critical frequency going to zero, suggesting that the technique of the rotating wire is well adapted to differentiate liquid- from gel-like behaviors, as defined in rheology.

In the present paper, we exploit this concept to study the mechanical response of the cytoplasm of mouse fibroblast NIH/3T3 and human cancer HeLa cells. It is first demonstrated that the micron-size wires specifically fabricated for the study are non-toxic, enter spontaneously inside living cells and that they are not comprised in membrane-bound compartments. Actuated by a low frequency external field, the wires rotate at the same angular speed than the field, whereas at higher frequency their motion exhibits back-and-forth oscillations. Values for the static viscosity, elastic modulus and relaxation time of the cytoplasm are derived from a complete analysis of the wire temporal trajectories. It is concluded that the interior of fibroblasts and HeLa cells can be appropriately described as viscoelastic liquids.

# Results

**Magnetic wires synthesis** Crystalline iron oxide nanoparticles were synthesized by co-precipitation of iron(II) and iron(III) salts in alkaline aqueous media and by further oxidation of the magnetite ($Fe_3O_4$) into maghemite ($\gamma$-$Fe_2O_3$). The size and dispersity (ratio between standard deviation and average diameter) of the particles prepared were determined from transmission electron microscopy ($D_{TEM}$ = 13.2 nm, $s$ = 0.23, Fig. 1 and Supplementary Figure 1). Their crystallinity and structure was assessed by electron beam microdiffraction (Supplementary Figure 2). The magnetization curves at different volume fractions were obtained by vibrating sample magnetometry and adjusted using a paramagnetic model[39,40]. The adjustment provided a specific magnetization of $3.5 \times 10^5$ A m$^{-1}$ and a magnetic diameter of 10.7 nm (Supplementary Figure 3). Light scattering was used to measure the weight-average molecular weight ($M_w$ = $12 \times 10^6$ g mol$^{-1}$) and the hydrodynamic diameter ($D_H$ = 27 nm) for the bare particles[40]. Values of the electrophoretic mobility and zeta-potential were also derived ($\mu_E = -3.8 \times 10^{-4}$ cm$^2$ V$^{-1}$, $\zeta$ = -48 mV, Supplementary Table 1) and indicated that the dispersions were stabilized by electrostatic forces[41].

For the wire synthesis, the particles were coated with poly(sodium acrylate)[39,41] (Aldrich) of molecular weight $M_W$ = 5100 g mol$^{-1}$. The wires were made according to a bottom-up co-assembly process using $\gamma$-$Fe_2O_3$ coated particles and oppositely charged polyelectrolytes. The polymer used was poly(trimethylammonium ethylacrylate)-*b*-poly(acrylamide) with molecular weights 11000 g mol$^{-1}$ for the cationic block and 30000 g mol$^{-1}$ for the neutral block[42,43]. Supplementary Figure 4 illustrates the protocol for the fabrication of the wires. Fig. 1b display a TEM image of a single wire illustrating its nanostructure, whereas Fig. 1c shows a collection of wires deposited on a glass substrate and observed by optical microscopy (×40). The as-prepared samples contained anisotropic objects of median length 12.2 μm and of diameter 0.49 μm (Fig. 1d). Prior to incubation with cells, the wires were sonicated during 120 s to shorten their lengths and sorted by magnetic separation. The sonicated wires had an average length of 2.4 μm and a dispersity of 0.35. The wires dispersion was then autoclaved (Tuttnauer Steam Sterilizer 2340M) at 120 °C and atmospheric pressure during 20 mn to prevent bacterial contamination during cell culture, concentrated by magnetic sedimentation at $10^6$ wires per μL and stored at 4 °C in a secure environment before use.

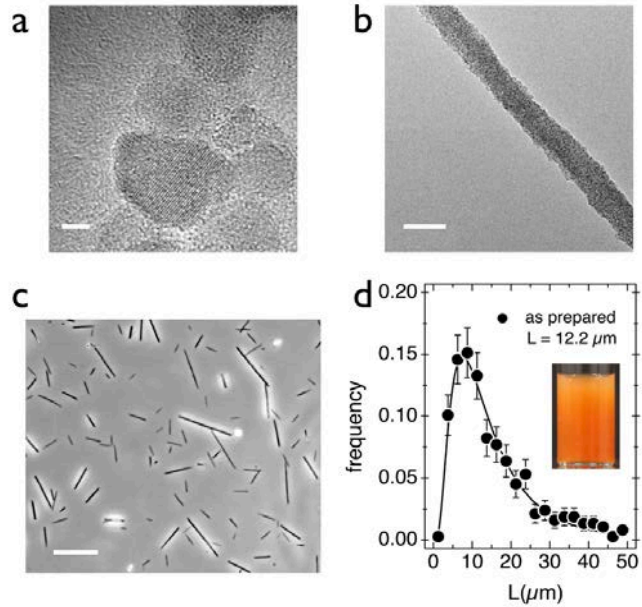

*Figure 1: Characterization of particles and wires*
*a) High-resolution transmission electron microscopy (TEM) of 13.2 nm iron oxide nanoparticles (bar 2 nm). b) TEM of magnetic wires made by co-assembly using oppositely charged particles and polyelectrolytes (bar 200 nm). c) Phase-contrast image of magnetic wires deposited on a glass substrate and observed by optical microscopy (×60, bar 10 µm). d) Size distribution of the wires adjusted using a log-normal distribution function with length of 12.2 µm and dispersity of 0.70. Error bars are defined as Standard Error Measurements. Inset: image of a vial containing a dispersion of wires at the concentration of $10^6$ wires µL$^{-1}$.*

**Wire microrheology technique tested on model fluids and gels** First we derive the equation of motion of a wire immersed in well-known rheological models[1] (Supplementary Figure 5) and submitted to a rotating magnetic field, and then assess the validity of the technique by comparing the predictions to observations obtained on a wormlike surfactant solution and on a polysaccharide gel.

*Newtonian liquid* - For a viscous liquid of viscosity $\eta_0$, a wire submitted to a rotating field experiences restoring a viscous torque that slows down its rotation. As a result, its motion undergoes a transition between a synchronous and an asynchronous rotation. The critical frequency $\omega_C$ between these two regimes reads:

$$\omega_C = \frac{3}{8}\frac{\mu_0 \Delta \chi}{\eta_0} g\left(\frac{L}{D}\right)\frac{D^2}{L^2} H^2 \qquad (1)$$

where $\mu_0$ is the permeability in vacuum, $L$ and $D$ the length and diameter of the wire, $H$ the amplitude of the magnetic excitation and $g\left(\frac{L}{D}\right) = Ln\left(\frac{L}{D}\right) - 0.662 + 0.917\frac{D}{L} - 0.050\left(\frac{D}{L}\right)^2$ is a dimensionless function of the anisotropy ratio[44,45]. In Eq. 1, $\Delta \chi =$ $\chi^2/(2+\chi)$ where $\chi$ denotes the magnetic susceptibility[27]. For data treatment, the geometrical characteristics are combined into the dimensionless parameter $L^* = L/D\sqrt{g(L/D)}$. Under these conditions, Eq. 1 becomes:

$$\omega_C = \frac{3\mu_0 \Delta\chi}{8\eta_0} \frac{H^2}{L^{*2}} \qquad (2)$$

The average angular velocity $\Omega(\omega)$ in the two regimes expresses as[27,35]:

$$\begin{aligned} \omega \leq \omega_C & \quad \Omega(\omega) = \omega \\ \omega \geq \omega_C & \quad \Omega(\omega) = \omega - \sqrt{\omega^2 - \omega_C^2} \end{aligned} \qquad (3)$$

With increasing frequency, the average velocity increases linearly, passes through a cusp-like maximum at the critical frequency and then decreases. The transition between the synchronous and asynchronous regimes was used to calibrate the wire rheometer and determine the susceptibility parameter $\Delta\chi$. Experiments were performed at T = 26 °C on a 85 wt. % water-glycerol mixture of static viscosity $\eta_0$ = 0.062 Pa s$^{-1}$. In a purely viscous fluid, the critical frequency is found to decreases as $L^{*-2}$ in agreement with Eq. 2. For wires made from 13.2 nm particles and block copolymers, we found $\Delta\chi = 2.3 \pm 0.7$, and a magnetic susceptibility $\chi = 3.6 \pm 0.9$ (Supplementary Figure 6).

*Maxwell fluid* – a Maxwell fluid is described by a spring and dashpot in series[1]. An actuated wire immersed in such a medium experiences a viscous and an elastic torque that both oppose the applied magnetic torque. The differential equation describing the wire motion has been derived and solved, leading to the following predictions (Supplementary Note 1). With increasing ω, the wire undergoes the same type of transition as the one described previously and the critical frequency $\omega_C$ expresses as in Eq. 1. The static viscosity $\eta_0$ in Eq. 1 is however replaced by the product $G\tau$, where $G$ and $\tau$ denote the shear elastic modulus and the relaxation time of the fluid. The set of equations in Eq. 3 is also identical to that of a Newtonian fluid. From the amplitude of the oscillations $\theta_B(\omega)$ (more precisely the angle by which the wire returns back after a period of increase) in the asynchronous regime, it is possible to determine the shear elastic modulus $G$ using[25]:

$$\lim_{\omega \to \infty} \theta_B(\omega) = \frac{3}{4} \frac{\mu_0 \Delta\chi}{G} g\left(\frac{L}{D}\right) \frac{D^2}{L^2} H^2 \qquad (4)$$

The above predictions were tested by monitoring the wire motion in a wormlike micellar solution made from cetylpyridinium chloride and sodium salicylate ([NaCl] = 0.5 M) at 2 wt. %[46]. The rheological parameters of the surfactant solution as determined by cone-and-plate rheometry were $\eta_0 = 1.0 \pm 0.1$ Pa s, $G = 7.1 \pm 0.1$ Pa and $\tau = 0.14 \pm 0.01$ s (Supplementary Note 2). From the evolution of the average angular velocity $\Omega(\omega)$ of actuated wires and from the position of the maximum (Eq. 3), a static viscosity of $1.3 \pm 0.3$ Pa s was obtained. From the oscillation amplitudes in the regime $\omega\tau \gg 1$ (Eq. 4), a modulus of $9.4 \pm 2$ Pa was derived (Supplementary Figure 7). Experiments performed with wires of different lengths and in various magnetic field conditions confirmed the good agreement with cone-and-plate rheometry, and demonstrates the ability of the technique to measure the linear viscoelasticity of a Maxwell fluid[25].

*Kelvin-Voigt solid* - a Kelvin-Voigt element aims to describe a viscoelastic solid, and it is represented by a spring and dashpot in parallel[1]. The resolution of the differential equation of motion of a wire leads to the following predictions (Supplementary Note 1). At all frequencies, the wire rotation exhibits back-and-forth oscillations at a frequency twice that of the field, and the average angular velocity $\Omega(\omega) = 0$. In addition, the amplitude of the oscillations $\theta_B(\omega)$ varies inversely with the shear elastic modulus according to:

$$\theta_B(\omega) = \frac{3}{4} \frac{\mu_0 \Delta \chi}{G'(\omega)} g(\frac{L}{D}) \frac{D^2}{L^2} H^2 \quad (5)$$

Gellan gum (phytagel, Sigma-Aldrich), a linear anionic polysaccharide comprising glucose, glucuronic acid and rhamnose building units[47] was added slowly to a 1 mM calcium chloride solution at room temperature with rapid stirring before heating up to 50 °C. Samples prepared at concentrations 0.3, 0.5, 0.75, 1 and 2 wt. % and studied by cone-and-plate rheometry exhibited a gel-like behavior: the elastic modulus exhibits scaling of the form $G'(\omega) \sim \omega^{0.2}$ and $G'(\omega) > G''(\omega)$ (Supplementary Note 3). At frequencies $5 \times 10^{-3}$ to 10 rad s$^{-1}$ the rotation angle of the wires reveal steady oscillations, and an average angular speed $\Omega(\omega)$ equal to zero, in agreement with the Kelvin-Voigt predictions (Supplementary Figure 8). Using Eq. 5, we found an elastic modulus $G'(\omega)$ of 2.5 ± 0.8 Pa, 9.5 ± 3 Pa and 95 ± 30 Pa for the 0.3, 0.5 and 1 wt. % samples, respectively, again in good agreement with the cone-and-plate data ($G'(\omega) = 3 \pm 0.5$ Pa, 13 ± 2 Pa and 130 ± 10 Pa). In conclusion of this part, we have found that theory correctly predicts the motion of rotating magnetic wires immersed in viscoelastic model systems and that the viscoelastic parameters retrieved, such as the viscosity or the elastic modulus are those of the linear shear rheology.

**Interactions with cells: wires are dispersed in the cytosol**  Here we establish the experimental conditions under which the wires interact with the cells, and in which part of the cytoplasm they are located. The localization of the wires inside the cells was determined by transmission electron microscopy. The incubation time was 24 h and the amount of wires per cell 10 for the two cell lines. Under such conditions, no perturbations of the cell morphology or cell cycle were observed (Supplementary Figure 9). In a previous report, it was shown that the capture rate for neutral wires was approximately 10% at 24 h, leading to an average number of one wire per cell in the present experiment[48]. Figs. 2a and 2b display TEM images of NIH/3T3 and HeLa cell interiors respectively. The elongated objects appearing in the TEM images are in the range 200 nm – 1 µm and they are made of densely packed particles of high electronic contrast. These threads are pieces of wires that were uptaken by the cells. horter than their initial lengths, the threads have also sharp and diffuse extremities. These observations are interpreted by assuming that elongated structures not in the plane of the microtome section were shortened during the sample preparation, an outcome that was already noted in previous studies[48,49]. In addition to wires, clusters in the range of 200 nm are also visible. These clusters result from the wire degradation occurring during incubation or the sample preparation. Close-up views of the TEM images in Fig. 2c and 2d indicate that the wires are not in membrane bound compartments[50]. This result was already pointed out in our previous study on fibroblasts. Here however, the TEM outcomes are similar for the cancerous HeLa cells and for human lymphoblasts (data not shown), indicating that the absence of endosomal membrane around internalized wires may be specific to anisotropic objects. Concerning the entry mechanism, several scenarios were suggested, including the perforation of the cellular membrane[51,52], macropinocytosis[53] or multiple combined processes[48,54]. At this stage, no definite conclusion can be made concerning the entry mechanism into NIH/3T3 or HeLa cells. The absence of cytoplasmic membrane around the internalized probes ensures however that the wires will probe the mechanical properties of the cytosol.

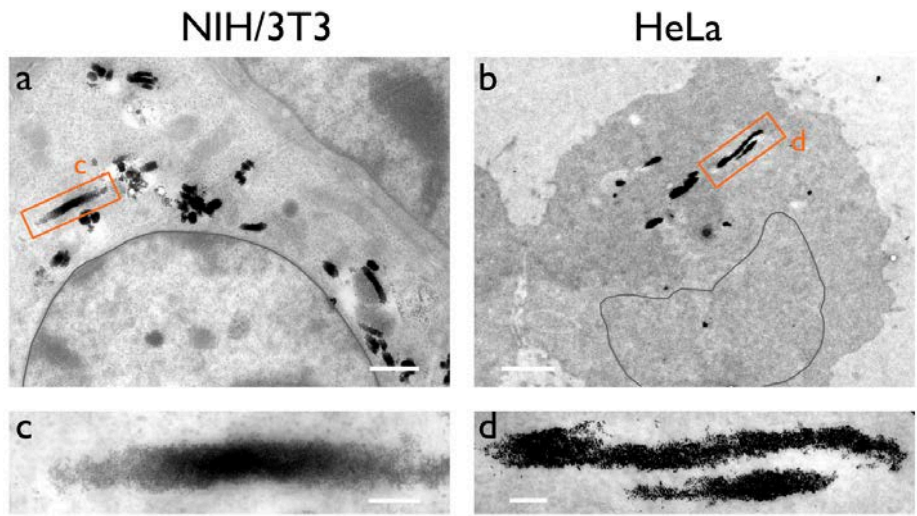

*Figure 2: Localization of wires inside living cells*
*Representative transmission electron microscopy image of a) NIH/3T3 fibroblasts and b) HeLa cancer cells incubated with 2.4 µm wires for 24 h at the concentration of 10 wires per cell. For the fibroblasts, clusters in the range of 200 nm are also visible. Scale bars in a) and b) are 1 and 2 µm respectively. c) and d) Close-up views of regions marked by an orange rectangle in Fig. 2a and 2b, respectively, indicating that the wires are not enclosed in membrane-bound compartments. Scale bars are 200 nm.*

**Time-resolved response under steady actuation in living cells**  We then examine the motions of wires internalized into cells and submitted to a 14 mT rotating field (see movies in Supplementary Information). In this part, emphasis is put on the time dependent behavior. Wires of lengths 2 to 6 µm were studied as a function of the frequencies $\omega$. For each condition, a 100 s movie was recorded and digitalized, from which the position of the center-of-mass and orientation angle of the wire were retrieved and plotted as a function of time. Figs. 3a show the rotation of a 2.8 µm wire internalized into a fibroblast at the angular frequency of 0.14 rad s$^{-1}$ and Fig. 3b illustrates the time dependence of the angle $\theta(t)$ at different frequencies. For the conditions tested, $\theta(t)$ increases linearly with time, and the slope $\Omega = \langle d\theta(t)/dt \rangle_t$ corresponds exactly to the actuating frequency. Here, $\langle ... \rangle_t$ denotes the time average in the steady regime. In this first regime, the wire rotates with the field, and $\theta(t) = \omega t$.

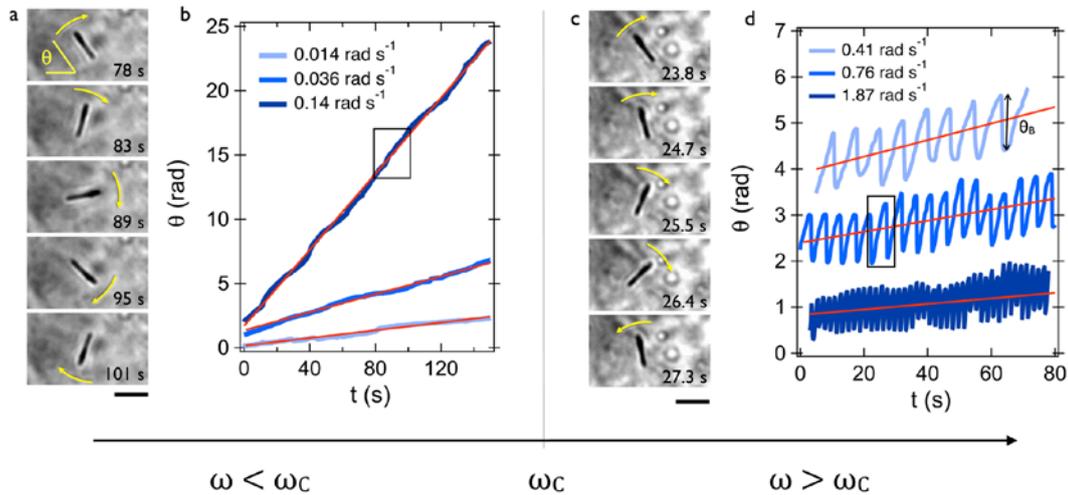

*Figure 3: Evidence of the wire rotational instability in living cells*
*a) Optical microscopy images of a 2.8 µm internalized wire subjected to a rotating field of 14 mT and the frequency of 0.14 rad s$^{-1}$. b) Time dependences of the angle $\theta(t)$ at varying frequencies $\omega$ = 0.014, 0.036 and 0.14 rad s$^{-1}$. The rectangle around 80 s provides the time range for the images on the left hand side. The straight lines in red were calculated from the expression $\theta(t) = \omega t$, indicating that the wire rotates synchronously with the field, and at the same angular frequency. c) and d) Same representation as in a) and b) but for angular frequencies (0.41, 0.76 and 1.87 rad s$^{-1}$) above the critical frequency $\omega_C$ = 0.15 rad s$^{-1}$. The microscopy images on the left-hand side shows that after a clockwise rotation, the wire comes back rapidly in an anti-clockwise motion, indicating that the wire rotation is hindered. Scale bars in a) and c) are 2 µm.*

Figs. 3c and 3d illustrates the change of regime as the angular frequency is increased above a critical value, here $\omega_C$ = 0.15 rad s$^{-1}$. The microscopy images on the left hand side show that after a clockwise steady rotation (between 23.8 and 26.4 s), the wire comes back rapidly by 50 degrees in an anti-clockwise motion, indicating that the wire rotation is hindered. On a longer period, the wires are animated of back-and-forth motion characteristic of the asynchronous regime (Fig. 3d). The traces recorded at $\omega$ = 0.41, 0.76 and 1.87 rad s$^{-1}$ display steady oscillations, with a rotation angle that continues to increase with time. In this second regime however, it is observed that $\Omega(\omega) \ll \omega$. The data shown in the figure are representative of the overall behavior found for wires in living fibroblast and HeLa cells. In these time resolved experiments, it was verified that the center-of-mass of the wires remained punctual during the steady or hindered rotations, insuring that the same volume of the cytosol is probed during the measurements. In conclusion to this part, with increasing frequency, wires dispersed in the intracellular medium of fibroblasts or of Hela cells undergo a transition between a synchronous and asynchronous regime. The critical frequency is found in the range 0.01 - 1 rad s$^{-1}$.

**Frequency dependence of the average rotation frequency in cells**  To elucidate the rheological profile of NIH/3T3 and HeLa intracellular medium, the $\theta(t)$-traces of internalized wires were analyzed and translated into a set of two parameters: the average rotation velocity $\Omega(\omega)$ and the amplitude $\theta_B(\omega) = \langle \theta_B(t,\omega) \rangle_t$ of the oscillations in the instable regime. As mentioned in the section on the validity of the magnetic wire spectroscopy, it was found that for Newton, Maxwell and Kelvin-Voigt models, $\Omega(\omega)$ and $\theta_B(\omega)$ display specific asymptotic behaviors as a function of the frequency[25]. For the viscous and viscoelastic liquids, the average frequency $\Omega(\omega)$ is superimposed and show a cusp-like maximum at $\omega_C$ whereas for a gel the average rotation velocity is constant and equal to 0. These different behaviors are illustrated in Fig. 4a.

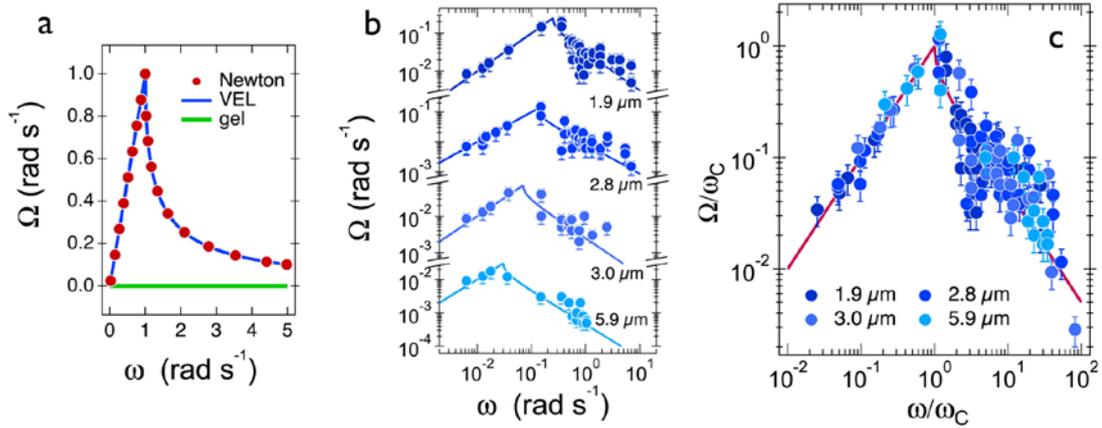

*Figure 4: Master curve for the wire average rotation velocity in cells*
*a) Average rotation velocity $\Omega(\omega)$ of a magnetic wire calculated for a Newtonian fluid, for a viscoelastic liquid (indicated as VEL) and for a gel according to a set of models listed in Supplementary Note 1. For the sake of convenience, the viscous and the viscoelastic liquids have here the same static viscosity and $\omega_C$ = 1 rad s$^{-1}$. b) $\Omega(\omega)$-evolution for wires of lengths between 1.9 µm and 5.9 µm internalized in the cytoplasm of mouse fibroblasts. The continuous lines (from Eq. 1) indicate the existence of a critical frequency in the rotation dynamics. c) Data from Fig. 4b plotted in reduced units, $\Omega/\omega_C$ versus $\omega/\omega_C$. The continuous lines are $\Omega(X) = X$ and $\Omega(X) = X - \sqrt{X^2 - 1^2}$ for $X \leq 1$ and $X \geq 1$, respectively. Error bars in b) and c) are defined as Standard Error Measurements.*

Fig. 4b displays the $\Omega(\omega)$-evolution for wires of lengths between 1.9 µm and 5.9 µm. With increasing frequency, the average velocity increases linearly, passes through a maximum at $\omega_C$ and then decreases. As already mentioned, the transition corresponds to the change of rotation regime, from synchronous rotation to back-and-forth oscillations. Note that the cut-off frequency increases as the length of the wire decreases (Fig. 4b). The data in Figs. 4 were adjusted using Eq. 3 [27,35]. In both regimes, the agreement between the data and the model is excellent. At high frequency, the $\Omega(\omega)$-data exhibit some scattering that could come from time-dependent effects of the viscosity, and be related to the cellular activity. Table I provides the values of the critical frequency and static viscosity for the experiments in Fig. 4b. There, $\omega_C$ varies from 0.03 to 0.2 rad s$^{-1}$, and is associated to static viscosities between 20 and 80 Pa s. Fig. 4c displays the data of Fig. 4b in reduced units, $\Omega/\omega_C$ *versus* $\omega/\omega_C$. In this representation, the $\Omega/\omega_C$-data are found to collapse onto the same master curve, now observed over four decades in $\omega$. The agreement obtained over a large frequency window attests to the reliability of the model. A comparison with the predictions shown in Fig. 4a suggests that the rheology of the intracellular medium is that of a viscous or of a viscoelastic liquid. The data in Fig. 4c also rule out the hypothesis of a gel-like rheology.

| Wire length | $\omega_C$<br>rad s$^{-1}$ | $\eta_0$<br>Pa s | $\lim_{\omega \to \infty} \theta_B(\omega)$<br>rad | $G$<br>Pa |
|---|---|---|---|---|
| $L$ = 1.9 μm | 0.20 ± 0.05 | 32 ± 8 | 0.8 ± 0.1 | 16 ± 2 |
| $L$ = 2.8 μm | 0.15 ± 0.02 | 26 ± 5 | 0.8 ± 0.1 | 9 ± 2 |
| $L$ = 3.0 μm | 0.07 ± 0.02 | 59 ± 15 | 0.8 ± 0.1 | 11 ± 2 |
| $L$ = 5.9 μm | 0.03 ± 0.01 | 78 ± 20 | 0.6 ± 0.1 | 8 ± 2 |

*Table I: Viscoelastic parameters of NIH/3T3 fibroblast*
*$\omega_C$ denotes the critical cut-off frequency between the synchronous and asynchronous regimes for the data in Fig. 4b. $\eta_0$ is the static viscosity derived from Eq. 1, $\lim_{\omega \to \infty} \theta_B(\omega)$ is the angle by which the wire returns back at high frequency and $G$ the elastic modulus obtained from Eq. 4.*

**Frequency dependence of the oscillation amplitude in living cells** Fig. 5a displays the behaviors of the oscillation amplitudes for Newtonian and Maxwell fluids as a function of the reduced frequency $\omega/\omega_C$. $\theta_B$ being related to the asynchronous regime, it is defined only for $\omega/\omega_C \geq 1$. For the viscoelastic predictions (labeled VEL), the curves were calculated for different relaxation times ranging from $0.01/\omega_C$ to $0.4/\omega_C$, as indicated in the margin. For the Newtonian fluid, the amplitude of the oscillations decreases with increasing frequency. In contrast, the Maxwell fluid displays a crossover between a viscous and an elastic regime that occurs at a fixed value of the reduced frequency, here $\omega\tau = 1/2$. The value of $1/2$ is explained by the fact that the frequency of the oscillations in this regime is twice that of the field. The existence of a plateau at high frequency is related to the elastic response of the fluid, and its height is given by Eq. 4[21].

Fig. 5b shows the dependence of $\theta_B$ in NIH/3T3 fibroblasts as a function of the reduced frequency for different wires of lengths between 1.9 and 5.9 μm. The major result that emerges from the figure is that the angles $\theta_B$ collected over many experimental conditions (*i.e.* changing $\omega$ and $L$) are well superimposed when plotted against the reduced frequency. Starting at $\theta_B$ = 1.5 rad slightly above $\omega_C$, the angle exhibits a moderate but steady decrease with increasing $\omega$. Also displayed in the figure is the prediction for a Newtonian fluid, which decrease is much stronger and cannot account for the data found in cells. Attempts to use a single-mode Maxwell model (Fig. 5a) were also inconclusive, and the $\theta_B$-data could not be reproduced. The adjustment was noticeably improved using a Generalized Maxwell Model (noted GMM on the figure) and assuming a constant distribution of relaxation times. The spread of the distribution was found to go from $1/\omega_C$ to $100/\omega_C$. For $\omega_C$-values of 0.1 rad s$^{-1}$, the time distribution extends hence from 0.1 to 10 s. From this fitting, the elastic modulus $G$ of the cell interior was estimated. We found an elastic modulus $G$ = 12 Pa (with a standard deviation of 4 Pa, $n$ = 9) for the fibroblasts and $G$ = 14 Pa (with a standard deviation of 9 Pa, $n$ = 10) for the HeLa cells. Such values for the elasticity are in agreement with the results obtained by tracking techniques, specifically those probing the interior of the cell[13]. In contrast, $G$-values found here are one to two decades smaller than that determined by techniques probing the mechanical response of the entire cell, or that of the cell surface[4,5,9]. Such differences were already discussed in the literature and for the latter experiments they were attributed to the cortical actin network[12].

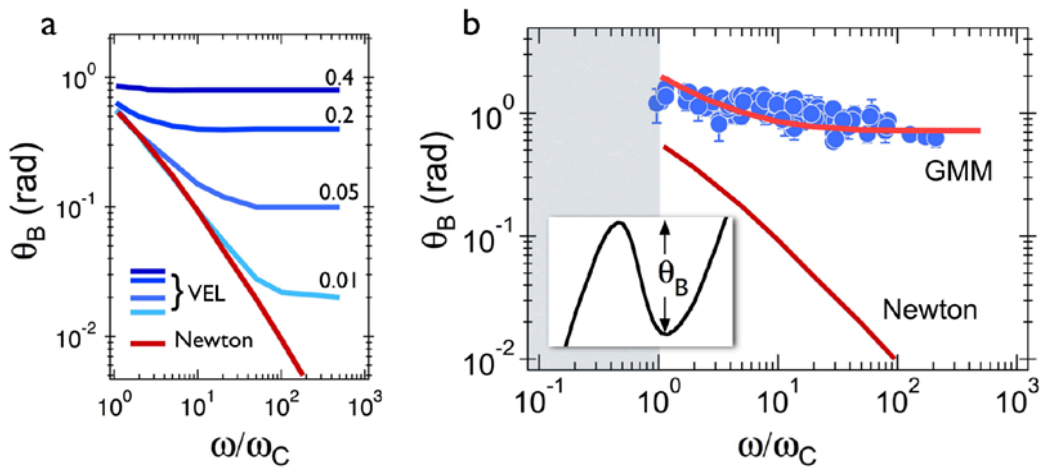

*Figure 5: Master curve for wire oscillation amplitude in cells*
**a)** Average rotation velocity $\Omega(\omega)$ of a magnetic wire calculated for a Newtonian fluid and for a viscoelastic liquid (indicated as VEL) of relaxation times ranging from $0.01/\omega_C$ (light blue) to $0.4/\omega_C$ (dark blue). **b)** Variation of $\theta_B$ as a function of the reduced frequency $\omega/\omega_C$ for wires of lengths 1.9 to 5.9 μm. The continuous lines were obtained from the Newtonian and from the Generalized Maxwell (GMM) models. Inset: definition of $\theta_B$. Error bars are defined as Standard Error Measurements.

## Discussion

The model of rotating wires in a viscoelastic liquid contains an important prediction: the critical frequency $\omega_C$ varies inversely with the square of the wire length. This relationship was verified for fluids of known viscosity[25,27] and tested again here for a 85% glycerol/water mixture (Supplementary Figure 6). Fig. 6a and 6b display the dependences of the critical frequency $\omega_C$ as a function of the reduced wire length $L^* = L/D\sqrt{g(L/D)}$ for the two cell lines studied. Least square calculations using a power law dependence of the form $\omega_C(L^*) \sim L^{*\alpha}$ provide exponents $\alpha$ = -3.0 ± 0.5 and $\alpha$ = -6.5 ± 1.0 for NIH/3T3 and HeLa, respectively (straight lines in red in Figs. 6). Two related effects could account for the discrepancy with respect to Eq. 2: the heterogeneities of the intracellular medium, which also contribute to a relatively broad scattering of the data points, and the possible dependence of the mechanical response on the length scale. This later feature was observed in active gels of F-actin using two-particle microrheology experiments[55]. By forcing the exponent to be -2 and repeating the fitting procedure, the static shear viscosity of fibroblasts could be estimated, providing $\eta_0$ = 47(+26/-14) Pa s. This value is larger than those reported using passive microrheology in different culture and probe conditions[13,21]. For the Hela cells, the deviation between the observed and predicted scaling is significant, and $\eta_0$ could not be determined. This result confirms the high sampling variability of rheological parameters associated with this cell line.

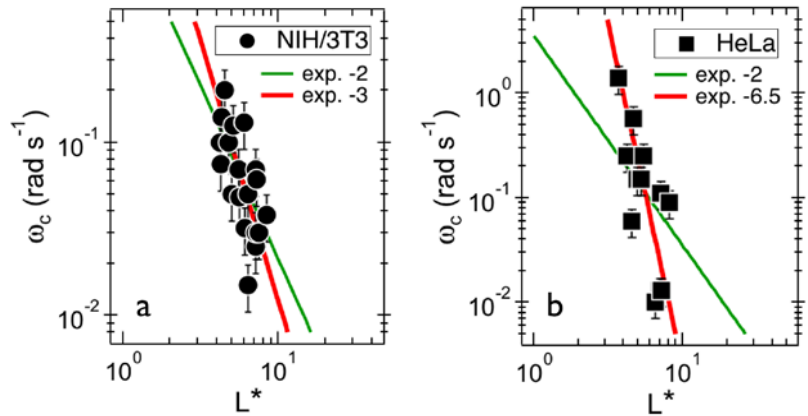

*Figure 6: Scaling critical frequency and wire length*
*Variation of the critical frequency $\omega_C$ as a function of the reduced wire length $L^* = L/D\sqrt{g(L/D)}$ for a) NIH/3T3 mouse fibroblasts and b) for HeLa cervical cancer cells. Straight lines in red are least-square fits using power laws with exponent -3 and -6.5 respectively, whereas straight lines in green are obtained from Eq. 2.*

In this work, we used magnetic wires between 2 and 6 µm to measure the static viscosity of living cells. The technique, known as rotational magnetic spectroscopy, is based on the experimental determination of a critical frequency $\omega_C$ between two different rotation regimes, one where the wire and the field are synchronous, and one where the wire perform back-and-forth oscillations. The measure of the viscosity is obtained directly from the critical frequency $\omega_C$, through the relationship given by Eq. 1. This wire-based technique is known for more than 20 years but was applied only recently to complex fluids composed of colloids of synthetic and natural origin[25,27,29,30,37,38,56]. To the best of our knowledge, it was not applied yet to living cells. In a first step, we assess the validity of the wire-based microrheology technique using known rheological models. We have found that theory and experiments agree well with each other, and that the viscoelastic parameters retrieved (including the static viscosity $\eta_0$ and the elastic modulus $G$) are those of the linear shear rheology. With living cells, one of the key results here is that the rotational instability as in the range $10^{-2} - 1$ rad s$^{-1}$ for the wires of magnetic susceptibility $\chi = 3.6$. It is therefore accessible to standard rotating field devices mounted on an optical microscopy platform. In this paper, the analysis has been extended to the amplitude of the oscillations in the asynchronous regime and to its frequency dependence and scaling. The high frequency limit was used to estimate the shear elastic modulus $G$ of the intracellular medium for the NIH/3T3 fibroblasts and for the HeLa cervical cancer cells. The modulus is in the range 5 - 20 Pa.

The determinations of the viscosity and elasticity confirm the viscoelastic character of the cytoplasm. However here, and in stark contrast to several studies[9,10,56], we conclude that the interior of living cells is best described in terms of a viscoelastic liquid, and not of an elastic gel. The conclusion was made possible by gaining access to a frequency range not explored before (down to $6 \times 10^{-3}$ rad s$^{-1}$), and for which the cytoplasm was shown to flow. The present magnetic rotational spectroscopy method does not allow to determine the distribution of relaxation times with accuracy, in part because of the restricted range of frequencies explored. An estimate is nevertheless provided and it is found that the time distribution is constant in the range 0.1 – 10 s[9]. The rheological model used to fit the data is in agreement with a Generalized Maxwell model combined with a distribution of relaxation times. In conclusion, the present work shows the potential of the wire-based magnetic rotation spectroscopy as an accurate rheological technique to distinguish between flow and yield stress behaviors in highly confined environment.

# Methods

**Microrheology set-up**   Bright field microscopy was used to monitor the actuation of the wires as a function of time. Stacks of images were acquired on an IX73 inverted microscope (Olympus) equipped with a 100× objective. For the measure of the magnetic properties of the wires, 65 µl of a dispersion of a know viscosity were deposited on a glass plate and sealed into to a Gene Frame® (Abgene/Advanced Biotech) dual adhesive system. The glass plate was introduced into a homemade device (Supplementary Figure 10) generating a rotational magnetic field, thanks to two pairs of coils working with a 90°-phase shift. An electronic set-up allowed measurements in the frequency range $10^{-2}$ - 100 rad s$^{-1}$ and at magnetic fields B = 0 – 20 mTesla. A stream of air directed toward the measuring cell was used to thermalize the sample at T = 26 or 37 °C. The image acquisition system consisted of an EXi Blue CCD camera (QImaging) working with Metaview (Universal Imaging Inc.). Images of wires were digitized and treated by the ImageJ software and plugins (http://rsbweb.nih.gov/ij/). The 3D motion was extracted from their 2D projection according a procedure described previously[26,44].

**Calibration of the wire based rheometer**   To determine the susceptibility parameter $\Delta\chi$ in Eqs. 1 and 2, steady rotation experiments were carried out on a fluid of known viscosity. A 85 wt. % glycerol-water mixture of static viscosity $\eta_0 =$ 0.062 Pa s$^{-1}$ (T = 26 °C) was used as suspending medium. In a typical optical microscopy experiment, a wire was first selected, its length and diameter measured with the objective 100× and it was then submitted to rotation frequency sweep from 0.1 to 100 rad s$^{-1}$. For a statistically relevant sample, the protocol was applied on 44 wires of lengths 2 – 20 µm and at a magnetic field of 7 mT. In Supplementary Figure 10, $\omega_C$ is shown as a function of $L^*$. The critical frequency was found to decrease as $\omega_C \sim L^{*-2}$, in agreement with the prediction of Eq. 2. From the prefactor ($3\mu_0 \Delta\chi H^2 / 8\eta_0 = 590 \pm 180$ rad s$^{-1}$), we infer $\Delta\chi = 2.3 \pm 0.7$, and $\chi = 3.6 \pm 0.9$. Knowing $\Delta\chi$, a wire submitted to a rotating field will be subjected to a known magnetic torque, a condition that is a prerequisite for quantitative rotational microrheology.

**Cell culture**   NIH/3T3 fibroblast (ATCC®-CRL-1658) and HeLa (ATCC®-CCL-2) cells were grown in T25-flasks as a monolayer in DMEM with high glucose (4.5 g L$^{-1}$) and stable glutamine (PAA Laboratories GmbH, Austria). The medium was supplemented with 10% fetal bovine serum (FBS) and 1% penicillin/streptomycin (PAA Laboratories GmbH, Austria), referred to as cell culture medium. Exponentially growing cultures were maintained in a humidified atmosphere of 5% $CO_2$ and 95% air at 37 °C, and in these conditions the plating efficiency was 70 – 90% and the cell duplication time was 12 – 14 h. Cell cultures were passaged twice weekly using trypsin–EDTA (PAA, Austria) to detach the cells from their culture flasks and wells. The cells were pelleted by centrifugation at 1200 rpm for 5 min. The supernatant was removed and cell pellets were re-suspended in assay medium and counted using a Malassez counting chamber.

**Transmission Electron Microscopy**   NIH/3T3 fibroblast cells were seeded onto the 6-well plate. After a 24 h incubation with 4 µm wires, the excess medium was removed. The cells were washed in 0.2 M phosphate buffer (PBS) and fixed in 2% glutaraldehyde-phosphate buffer (0.1 M) for 1 h at room temperature. Fixed cells were further washed in 0.2 M PBS. The cells were then postfixed in a 1% osmium-phosphate buffer for 45 min at room temperature in dark conditions. After several washes with 0.2 M PBS, the samples were dehydrated by addition of ethanol. Samples were infiltrated in 1:1 ethanol:epon resin for 1 h and finally in 100% epon resin for 48 h at 60 °C for polymerization. 90 nm-thick sections were cut with an ultramicrotome (LEICA, Ultracut UCT) and picked up on copper-rhodium grids. They were then stained for 7 min in a 2%-uranyl acetate and for 7 min in a 0.2%-lead citrate. Grids were analyzed with a transmission electron microscope (ZEISS, EM 912 OMEGA) equipped with a LaB$_6$ filament, at 80 kV. Images were recorded with a digital camera (SS-CCD, Proscan 1024×1024), and the iTEM software.

## Acknowledgments
K. Anselme, A. Baeza, A. Cebers, L. Chevry, E. Fodor, M.-A. Fardin, A. Hallou, F. Montel and P. Visco are acknowledged for fruitful discussions. Interns who participated to the research, C. Leverrier, A. Conte-Daban, C. Lixi, L. Carvhalo and

R. Chan are also acknowledged. The author is grateful to the ImagoSeine facility (Jacques Monod Institute, Paris, France), and to France BioImaging infrastructure supported by the French National Research Agency (ANR-10-INSB-04, « Investments for the future »), and to L. Vitorazi, F. Mousseau and R. Le Borgne for the TEM images of the HeLa cells. The Laboratoire Physico-chimie des Electrolytes, Colloïdes et Sciences Analytiques (UMR Université Pierre et Marie Curie-CNRS n° 7612) is acknowledged for providing us with the magnetic nanoparticles.

## Author contributions

J.-F.B. conceived the project, developed, performed and supervised the experiments, carried out the data analysis and calculations, and wrote the paper.

## Additional information

**Supplementary Information** accompanies this paper at http://www.nature.com/naturecommunications

**Competing financial interests**: The author declares no competing financial interests.